\documentstyle [10pt] {article}
\def\be{\begin{equation}}
\def\ee{\end{equation}}
\def\bea{\begin{eqnarray}}
\def\eea{\end{eqnarray}}

\begin{document}
%%%%%%%%%%%%%%%%%%%%%%%%%%%%%%%%%%%%%%%%%%%%%%%%%%%%%%%%%%%%%%%%%%%%
%\draft

\title{  On space-times admitting  shear-free, irrotational, 
geodesic null 
congruences}

\author{Alicia M. Sintes \\
{\small Max-Planck-Institut f\"ur Gravitationsphysik
(Albert-Einstein-Institut), }\\
{\small Schlaatzweg 1, 14473 Potsdam, Germany}\\ \\
Alan A. Coley and Des J. McManus \\
{\small Department of Mathematics, Statistics and
 Computing Science.} \\ {\small Dalhousie University. 
 Halifax, NS. Canada B3H 3J5}}
\date{}
 \maketitle

\begin{abstract}
Space-times admitting a shear-free, irrotational, geodesic 
null congruence are studied. Attention is focused on those 
space-times in which the gravitational field is a 
combination of a perfect fluid and null radiation.
 \end{abstract}
%\pacs{}

\section{Introduction}

In this article we wish to extend  earlier work on shear-free,
irrotational and geodesic (SIG) timelike and spacelike 
congruences \cite{des,des2}
to SIG {\it null}
 congruences. The fact that we are dealing 
with null congruences means that we have to approach the problem in a 
completely different way; we must make extensive use of the Newman-Penrose
formalism.  

Thus, we wish to study a congruence of curves whose tangent vector ${\bf k}$
is null and geodesic. Hence, we have a family of null geodesics
$x^a=x^a(y^{\alpha},v)$,
where $y^{\alpha}$ distinguishes the different geodesics, and $v$ is the
affine parameter along a fixed geodesic. The null tangent vector is
$k^a={\partial x^a \over \partial v} $,
and satisfies ${k^a}_{;b}k^b=0$.
The spin coefficients are defined in \cite{Kramer}, 
where $\rho=-(\theta + i\omega)$
is called the complex divergence and $\sigma$ is the complex shear.
The geodesic condition implies that the spin coefficient $\kappa$ vanishes
and
$\epsilon +\bar\epsilon =0$ follows from the choice of an affine parameter
along the congruence.
The congruence is said to be shear-free if
$\sigma=0$. 
Also, from the relation 
$k_{[a;b}k_{c]}=(\bar \rho-\rho)\bar m_{[a}m_bk_{c]}$  \cite{Pirani},
it follows that
 $w=0$ (i.e.,  zero twist) is a necessary and sufficient condition
for ${\bf k}$ to be hypersurface orthogonal.

First we shall  
  briefly review  some of the
 results of relevance to this work.
Goldberg and Sachs  \cite{Gold} proved
 that if a gravitational field contains
a shear-free, geodesic, null congruence ${\bf k}$, then
$\kappa=\sigma=0$,
and if
\be
R_{ab}k^ak^b=R_{ab}k^am^b=R_{ab}m^am^b=0 \ ,\label{s1}
\ee
then the field is algebraically special (i.e., $\Psi_0=\Psi_1=0$),
and ${\bf k}$ is a degenerate eigendirection. In addition, a vacuum metric is 
algebraically special if and only if it contains a shear-free geodesic
null congruence.

A space-time admits a geodesic, shear-free, twist-free 
($\kappa=\sigma=\omega=0$) and diverging ($\rho=\bar\rho=\theta=-1/r$)
null congruence ${\bf k}$, and satisfies (\ref{s1}),
if and only if the  metric can be written in the form
\be
ds^2=2r^2 P^{-2}(z,\bar z,u)dzd\bar z -2dudr -2H(z,\bar z,r,u)du^2 \ .
\ee
Robinson-Trautman models \cite{Robi} with this metric
 have been found for vacuum, Einstein-Maxwell
and pure radiation fields with or without a 
cosmological constant \cite{Kramer}.

For geodesic null vector fields we have that
$(\theta +i\omega)_{,a}k^a+ (\theta +i\omega)^2+\sigma\bar\sigma=
-R_{ab}k^ak^b/2$.
Therefore, in the non-diverging case (i.e., $\rho=-(\theta +i\omega)=0$),
if the 
 energy
condition $T_{ab}k^ak^b\ge 0$
is satisfied,
it follows that
  $\sigma=0=R_{ab}k^ak^b$.
Thus, non-twisting (and therefore geodesic) and non-expanding null
congruences must be shear-free. Hence, the space-time is algebraically special,
and it corresponds  to vacuum, Einstein-Maxwell, and pure radiation field.
Perfect fluid solutions violate $R_{ab}k^ak^b=0$ unless $\mu+p=0$. 
This class of solutions has been studied by Kundt \cite{Kundt}.

Another important case corresponds to the Kerr-Schild metric, which is
given by $g_{ab}=\eta_{ab}-2\phi k_ak_b$. The null vector ${\bf k}$ of a 
Kerr-Schild metric is geodesic if and only if the energy-momentum tensor 
obeys the condition $T_{ab}k^ak^b=0$, and then ${\bf k}$ is a multiple
principal null direction of the Weyl tensor and the space-time  is
algebraically special. The general properties of the Kerr-Schild metrics
and their applications to vacuum,  Einstein-Maxwell, and pure radiation
space-times can be found in \cite{Kramer}.

Finally, we note the algebraically special perfect fluid space-times
 corresponding to the generalized
Robinson-Trautman solutions
investigated by Wainwright \cite{Wain}. 
They are characterized by a multiple null 
eigenvector ${\bf k}$  of the Weyl tensor which is geodesic, 
shear-free, and twist-free but expanding (i.e., $\Psi_o=\Psi_1=0$, 
 $\kappa=\sigma=\omega=0$, $\rho=\bar\rho\not= 0$), 
and the four-velocity obeys $u_{[a;b}u_{c]}=0$, $k_{[c}k_{a];b}u^b=0$.
The line-element of the space-time can be written in the form
\be
ds^2= -{1\over 2}\chi^2(r,u)P^{-2}(z,\bar z, u)dzd\bar z +2du(dr-Udu) 
\ , \label{11}
\ee
with
\be
U=r(\ln P)_{,u}+ U^0(z,\bar z, u)+ S(r,u) \ , \quad \chi_{,r}\not =0
\ ,\quad  {\chi_{,rr}\over \chi}\le 0 \ .
\ee
In this case no dust solutions  nor solutions of Petrov types $III$ and $N$ are
possible. 

\section{Analysis}
Let us consider 
space-times admitting a shear-free, irrotational, geodesic null
congruence
 in which the source of the gravitational field is a 
{\it combination of a perfect fluid and null radiation}, 
so that  the energy-momentum tensor has the form
\be
T_{ab}=(\mu+p)u_au_b -p g_{ab} +\phi^2k_ak_b \ ,
\label{13}
\ee
where $u^a$ is the four-velocity of the fluid, 
$\mu$ and $p$ are the density and the pressure of
the fluid, respectively, and ${\bf k}$ is a null
vector. The null radiation is geodesic, twist-free, and shear-free, and defines 
the null congruence.
%%%%%%%%%%%%
%The energy momentum tensor (\ref{13}) 
%is equivalent to a single fluid with
%a non-zero heat flux and a non-zero anisotropic stress tensor.
%\be
%\hat T_{ab}=\bar\mu u_au_b +\bar p h_{ab} +u_aq_b +u_bq_a +\pi_{ab} \ .
%\label{14}
%\ee
%Comparing (\ref{13}) to (\ref{14}) we find: 
%$\bar\mu= \mu+ \phi^2 ({\bf k\cdot u})^2$,
% $\bar p= p + 1/3 \phi^2 ({\bf k\cdot u})^2$,
% $q_a=-\phi^2({\bf k\cdot u}) [k_a + ({\bf k\cdot u})u_a ]$ and
% $\pi_{ab}= \phi^2({\bf k\cdot u})^2u_au_b 
% - 1/3\phi^2({\bf k\cdot u})^2 h_{ab}
%+ \phi^2({\bf k\cdot u})(u_ak_b+u_bk_a) + \phi^2k_ak_b$.
%%%%%%%%%%%%%%%% 
 Wainwright  \cite{Wain} proved  that for a 
 space-time in which there exists a SIG
 null congruence, coordinates can be chosen so that the metric takes on the
 simplified form (\ref{11}) with $u=x^1$, $r=x^2$, $z=x^3+i x^4$, the 
 tangent field of the null congruence is given by
 $k^a=\delta^a_2$, $k_a=\delta^1_a$, and we can introduce the null tetrad
 \bea
k^a= \delta^a_r \ ,  & l^a= \delta^a_u+ U\delta^a_r \ , & 
m^a=P\chi^{-1}( \delta^a_3+ i\delta^a_4 ) \ , \\
k_a=\delta^u_a \ , & l_a= -U \delta^u_a + \delta^r_a  \ ,&
m_a=P^{-1}\chi (\delta^3_a + i \delta^4_a)/2 \ .
 \eea
With the sign convention used here  
 we have that $u^au_a=k^al_a=1=-m^a\bar m_a$. 
  Note that the null radiation is everywhere tangent to the repeated null
 congruence of the space-time.
 
 First, since $\Phi_{01}\equiv-{1\over 2}R_{ab}k^am^b=0$, we
 conclude that the four-velocity satisfies $u^am_a=0$, and hence it can be
 expressed in terms of the null tetrad by
 \be
 u^a={1\over \sqrt{2} B}(B^2 k^a +l^a) \quad {\rm and} \quad
 u_a={1\over \sqrt{2} B}[(B^2-U)\delta^u_a + \delta^r_a] \ , \label{36}
 \ee
 for some function $B$. The conditions 
 $\Phi_{02}\equiv-{1\over 2}R_{ab}m^am^b=0$ and 
 $\Phi_{12}\equiv-{1\over 2}R_{ab}m^al^b=0$ are satisfied identically.
 The non-zero components of the Ricci tensor are
 \bea 
 & &\Phi_{00}\equiv-{1\over 2}(R_{ab}-{1\over 4}Rg_{ab})k^ak^b=
 {1\over 2}(\mu+p)({\bf k\cdot u})^2 \ , \\
  & &\Phi_{11}\equiv-{1\over 4}(R_{ab}-{1\over 4}Rg_{ab})(k^al^b+m^a\bar m^b)=
 {1\over 4}(\mu+p)({\bf k\cdot u}) ({\bf l\cdot u})\ , \\
  & &\Phi_{22}\equiv-{1\over 2}(R_{ab}-{1\over 4}Rg_{ab})l^al^b=
 {1\over 2}(\mu+p)({\bf l\cdot u})^2 +{1\over 2}\phi^2 \ .
 \eea
In addition, since ${\bf k\cdot u}={1\over \sqrt{2} B}$ and
 ${\bf l\cdot u}={1\over \sqrt{2}} B$ implies 
 ${\bf l\cdot u}=B^2({\bf k\cdot u})$, we obtain
 \bea
 B^2\Phi_{00}&=& 2\Phi_{11} \  , \label{34}\\
 B^4\Phi_{00}&=&\Phi_{22}- {1\over 2}\phi^2  \ . \label{35}
 \eea
 
 If we now assume that the  fluid is non-rotating, then
 $ B^2=U+F(r,u)$,
and the compatibility condition (\ref{34}) can  be written as
 \be
 (U+F)\Phi_{00}=2\Phi_{11}\ .\label{39}
 \ee
 On differentiating this equation successively with respect to $z$ and $r$,
 we obtain the restriction 
\be
(\chi^2)_{,rrr}[{U^0}_{,z}+r(\ln P)_{,uz}]=0 \ .
\ee 
There are consequently 
 two different cases to consider.

In the first case  ${U^0}_{,z}+r(\ln P)_{,uz}=0$, which is
equivalent to ${U^0}_{,z}=(\ln P)_{,uz}=0$,
so that  
$P=P(z,\bar z)$ and $U^0=U^0(u)$. Obviously, the solutions admit a 
multiply transitive group of motions, $G_3$, acting on the 2-spaces $r=$const,
$u=$const, of constant curvature, and belong to class $II$ of 
Stewart and Ellis \cite{Stew}. The metric (\ref{11}) can then be rewritten as
\be
ds^2=-\chi^2(r,u){2dzd\bar z\over (1+{k\over 2}z\bar z)^2 }+
2du(dr-U(r,u)du) \ . \label{44}
\ee
The non-zero Ricci components are given by
\bea
& & \Phi_{00}=-{\chi_{,rr}\over \chi} \ , \\
& & \Phi_{11}={\chi_{,r}\chi_{,u}\over 2\chi^2} + 
{(\chi_{,r})^2U\over 2\chi^2}- {U_{,rr}\over 4} + {k\over 4 \chi^2} \ , \\
& & \Phi_{22}={\chi_{,u}U_{,r}\over \chi}- {\chi_{,uu}\over \chi}
-2{\chi_{,ur}U \over \chi}-{\chi_{,r}U_{,u}\over \chi}
-{\chi_{,rr}U^2 \over \chi} \ ,
\eea
and the Ricci scalar is given by
\be
{R\over 2}=12\Lambda= 4{\chi_{,r}U_{,r}\over \chi}+ 2{\chi_{,r}\chi_{,u}\over \chi^2}
+2 {(\chi_{,r})^2U\over \chi^2} + 4 {\chi_{,ur}\over \chi}+ U_{,rr}
+4 {\chi_{,rr}U \over \chi} +{k\over \chi^2} \ .
\ee
Hence, the metric (\ref{44}) can be interpreted as pure  radiation plus a
perfect fluid where $\mu$ and $p$ are given by
\be
\mu={R \over 4} +6\Phi_{11}\ , \quad p=-{R \over 4}+2\Phi_{11}\ ,
\label{45}
\ee
$u_a$ is determined by (\ref{36}) with $B^2=2 \Phi_{11}/\Phi_{00}$, and
$\phi^2$ is given by
\be
\phi^2=2\left( \Phi_{22}-4{\Phi_{11}^2\over \Phi_{00}}\right) \ .
\label{47}
\ee

 In the second case (i.e., ${\chi^2}_{,rrr}=0$) two possibilities arise:
 \bea
 (i) & \chi^2=\epsilon r,\qquad  & \epsilon= \pm 1 \label{52}\\
 (ii) & \quad \chi^2=\epsilon(r^2-k^2), & k=const \ .\label{53}
 \eea
In both subcases $\chi=\chi(r)$, 
and they can be written together as
$\chi^2=ar^2+2br+c $,
with $a$, $b$, $c$ taken to be appropriate constants. {}From equation (\ref{39})
we obtain
\be
aU^0 -b(\ln P)_{,u} + K =G(u) \ , \label{55}
\ee
and
\be
{1\over 2} [\chi^2 S_{,r} -S(\chi^2)_{,r}]_{,r} + {F \Sigma\over \chi^2}=
G(u) \ , \label{56}
\ee
where 
$K\equiv 4P^2 (\ln P)_{z\bar z}$, $\Sigma\equiv b^2-ac$,
and $G(u)$ is an arbitrary function of $u$. 

Subcase $(i)$: $a=c=0$, $b=\epsilon/2$. Integrating equation (\ref{56})
we see that $S$ can be written in the form
\be
S=rh(u)+2\epsilon G(u)r\ln r -f(u) 
-{1\over 2}r\int {dr\over r^2}\int^r {d\hat r\over \hat r}F(\hat r,u) \ ,
\label{58}
\ee 
where $h(u)$ and $f(u)$ are arbitrary functions of $u$.

Subcase $(ii)$: $a=\epsilon$, $b=0$, $c=-\epsilon k^2$, $\Sigma=k^2$.
We obtain
\be
S=-\epsilon G(u) + f(u) \chi^2\int {dr\over \chi^4} + 
h(u)\chi^2 -2k^2\chi^2\int{dr\over \chi^4}\int^r 
{d\hat r\over \chi^2(\hat r)}F(\hat r,u) \ .\label{59}
\ee

Therefore, the metric (\ref{11}) with $\chi(r)$ given by (\ref{52}) or
(\ref{53}), $S(r,u)$ given by  (\ref{58}) or (\ref{59}), and 
$P(z, \bar z, u)$ satisfying (\ref{55}) can be interpreted
as pure radiation plus  a perfect fluid, in which the four-velocity is
determined by (\ref{36}) and $\phi^2$, $\mu$ and $p$ are determined
by (\ref{45}) and (\ref{47}), respectively.

\section*{Acknowledgments}
This work was supported by the European Union, 
TMR Contract No. ERBFMBICT961479 (AMS), the Natural Sciences and
Engineering Research Council of Canada (AAC) and the Canadian 
Institute for Theoretical Astrophysics (DJM).

%\begin{references}


\begin{thebibliography}{99}
\bibitem{des} A.A. Coley and D.J. McManus, Class. Quantum Grav.,
 {\bf 11}(1994)1261.
\bibitem{des2}   D.J. McManus and A.A. Coley, Class. Quantum Grav.,
 {\bf 11}(1994)2045.
 
\bibitem{Kramer} D. Kramer, H. Stephani, M.A.H. MacCallum and E. Herlt, {\em
Exact Solutions of Einstein's Field Equations}, Deutscher Verlag der
Wissenschaften, Berlin (1980).
\bibitem{Pirani} F.A.E. Pirani, in {\em Lectures on General Relativity,
1964 Brandeis Summer Institute in Theoretical Physics}, Vol. 1. 
Prentice-Hall, Englewood Cliffs, NJ (1965).

 \bibitem{Gold} J.N. Goldberg and R.K. Sachs, Acta. Phys. Polon., Suppl.
 {\bf 22}(1962)13.
 \bibitem{Robi} I. Robinson and A. Trautman, Proc. Roy. Soc. Lond.,
  {\bf A265}(1962)463.
\bibitem{Kundt} W. Kundt, Z. Phys., {\bf 163}(1961)77.
 \bibitem{Wain} J. Wainwright,  Int. J. Theor. Phys., {\bf 10}(1974)39. 
 \bibitem{Stew} J.M. Stewart and G.F.R. Ellis, J. Math. Phys., 
 {\bf 9}(1968)1072. 
 
 %\end{references}
 \end{thebibliography}
\end{document}